\documentclass[runningheads]{llncs}
\usepackage{graphicx}
\usepackage{hyperref}

\usepackage{amsthm}
\begin{document}
\newtheorem{defi}[theorem]{Definition}
\newtheorem*{prel}{Preliminaries}
\title{G2GML: Graph to Graph Mapping Language for Bridging RDF and Property Graphs} 
\titlerunning{G2GML: Graph to Graph Mapping Language}
%
\author{Hirokazu Chiba\inst{1}\orcidID{0000-0003-4062-8903} \and Ryota Yamanaka\inst{2} \and Shota Matsumoto\inst{3}}
\authorrunning{H. Chiba {\itshape et al.}}

\setlength\abovecaptionskip{0ex}
%
\institute{
Database Center for Life Science, Chiba 277-0871, Japan\\
\email{chiba@dbcls.rois.ac.jp}
\and
Oracle Corporation, Bangkok 10500, Thailand\\
\email{ryota.yamanaka@oracle.com}
\and
Lifematics Inc., Tokyo 101-0041, Japan\\
\email{shota.matsumoto@lifematics.co.jp}
}
\maketitle
\begin{abstract}
How can we maximize the value of accumulated RDF data? Whereas the RDF data can be queried using the SPARQL language, even the SPARQL-based operation has a limitation in implementing traversal or analytical algorithms. Recently, a variety of database implementations dedicated to analyses on the property graph (PG) model have emerged. Importing RDF datasets into these graph analysis engines provides access to the accumulated datasets through various application interfaces. However, the RDF model and the PG model are not interoperable. Here, we developed a framework based on the Graph to Graph Mapping Language (G2GML) for mapping RDF graphs to PGs to make the most of accumulated RDF data. Using this framework, accumulated graph data described in the RDF model can be converted to the PG model, which can then be loaded to graph database engines for further analysis. For supporting different graph database implementations, we redefined the PG model and proposed its exchangeable serialization formats. We demonstrate several use cases, where publicly available RDF data are extracted and converted to PGs. This study bridges RDF and PGs and contributes to interoperable management of knowledge graphs, thereby expanding the use cases of accumulated RDF data.

\end{abstract}

\section{Introduction}

Increasing amounts of scientific and social data are being published in the form of the Resource Description Framework (RDF)~\cite{rdf}, which presently constitutes a large open data cloud. DBpedia~\cite{dbpedia} and Wikidata~\cite{wikidata} are well-known examples of such RDF datasets. SPARQL~\cite{sparql} is a protocol and query language that serves as a standardized interface for RDF data. This standardized data model and interface enables the construction of integrated graph data. However, the lack of an interface for graph-based analysis and performant traversal limits use cases of the graph data.

Recently, the property graph (PG) model~\cite{angles1,angles2} has been increasingly attracting attention in the context of graph analysis~\cite{agri}. Various graph database engines, including Neo4j~\cite{neo4j}, Oracle Database~\cite{oracle}, and Amazon Neptune~\cite{neptune} adopt this model. These graph database engines support algorithms for traversing or analyzing graphs. However, few datasets are published in the PG model and the lack of an ecosystem for exchanging data in the PG model limits the application of these powerful engines.

In light of this situation, developing a method to transform RDF into PG would be highly valuable. One of the practical issues faced by this challenge is the lack of a standardized PG model.
Another issue is that the transformation between RDF and PG is not straightforward due to the differences in their models. 
In RDF graphs, all information is expressed by triples (node-edge-node), whereas in PGs, arbitrary information can be contained in each node and edge as the key-value form. 
Although this issue was previously addressed on the basis of predefined transformations~\cite{hartig},
users still cannot control the mapping for their specific use cases.

In this study, we redefine the PG model incorporating the differences in existing models and propose serialization formats based on the data model. We further propose a graph to graph mapping framework based on the Graph to Graph Mapping Language (G2GML). Employing this mapping framework, accumulated graph data described in RDF can be converted into PGs, which can then be loaded into several graph database engines for further analysis. We demonstrate several use cases, where publicly available RDF data is extracted and converted to PGs. Thus, this study provides the foundation for the interoperability between knowledge graphs.

The main contributions of this study are as follows: 1) language design of G2GML and 2) its prototype implementation. Furthermore, we propose 3) the common PG model and its serialization, which are essential to ensure that this mapping framework is independent from the implementations of databases.

\section{Graph to Graph Mapping}

\subsection{Overview}

\begin{figure}
\center
\includegraphics[width=1.0\textwidth]{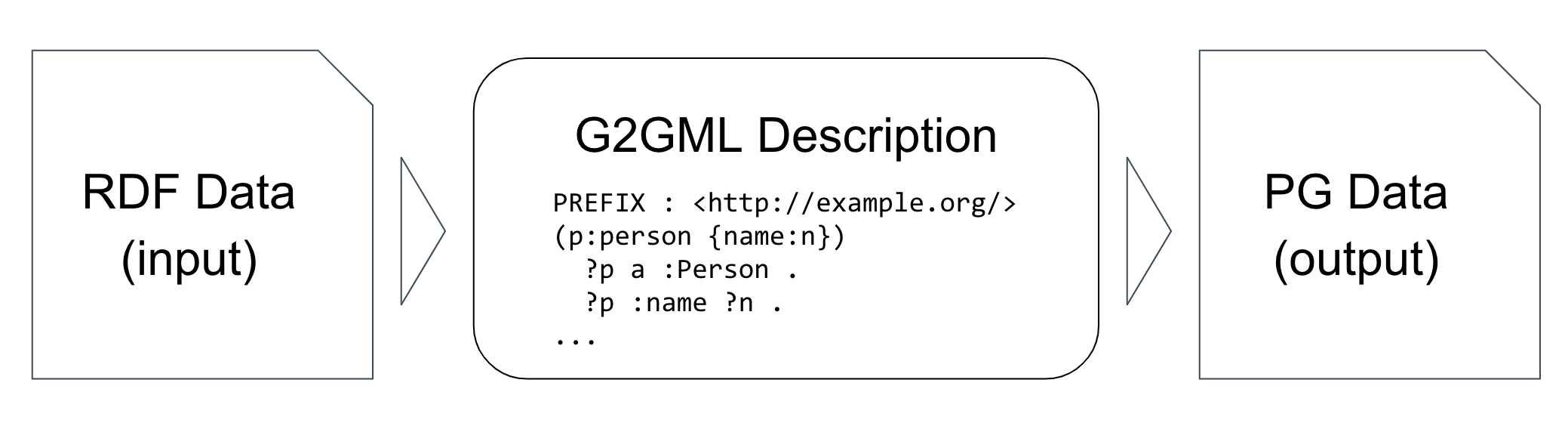}
\caption{Overview of mapping from RDF to PG}
\label{fig:dataflow}
\end{figure}

We provide an overview of the graph to graph mapping (G2G mapping) framework (Figure~\ref{fig:dataflow}).

In this framework, users describe mappings from RDF to PG in G2GML.
According to this G2GML description, the input RDF dataset is converted into a PG dataset. The new dataset can also be optionally saved in specific formats for loading into major graph database implementations.

G2GML is a declarative language comprising pairs of RDF patterns and PG patterns. 
The core concept of a G2GML description can be represented by a map from RDF subgraphs, which match specified SPARQL patterns, to PG components.

\subsection{G2GML}

G2GML is a language to describe maps from RDF graphs to PGs. It represents a domain-specific declarative language, where programmers only have to describe patterns of RDF and PG. The RDF pattern is described in terms of the SPARQL syntax, and the PG pattern is described in a syntax derived from openCypher \cite{openCypher}.

\begin{prel}
An RDF triple $(s, p, o)$ is an element of $(I \cup B) \times I \times (I \cup B \cup L)$, where $I$, $L$, and $B$ are a set of IRIs, a set of literals, and a set of blank nodes, respectively, which are considered pairwise disjoint. For an RDF triple $(s, p, o)$, $s$ is the subject, $p$ the predicate, and $o$ the object. An RDF graph is defined as a finite set of RDF triples.
\end{prel}

\par
A G2GML description comprises one or more \textit{node maps} and \textit{edge maps}.
A \textit{node map} is a function that maps resources from an RDF graph to a set of \textit{PG nodes}. 
Similarly, an \textit{edge map} is a function that maps resources from an RDF graph to a set of \textit{PG edges}.
Here, we define the syntax of G2GML as follows. 

\begin{defi}[EBNF notation of G2GML]
\label{defi:G2GML}
\leavevmode
\begin{verbatim}
 g2gml   ::= SPARQL_PREFIX* mapping+
 mapping ::= ( node_pat | edge_pat ) NEWLINE
               INDENTED_RDF_PATTERN NEWLINE
\end{verbatim}
\end{defi}
In Definition \ref{defi:G2GML}, \texttt{SPARQL\_PREFIX}
 and \texttt{INDENTED\_RDF\_PATTERN} are almost the same as \textit{PrefixDecl} and \textit{GroupGraphPattern} in SPARQL syntax (see EBNF notation of SPARQL grammar~\cite{sparql}).
The only difference is that \texttt{INDENTED\_RDF\_PATTERN} must have one or more white spaces at the beginning of each line.
The symbols \texttt{node\_pat} and \texttt{edge\_pat} are described according to Definition \ref{defi:pgnode} and \ref{defi:pgedge}. \texttt{NEWLINE} is used as an abstract symbol corresponding to the appropriate character to start a new line. Minimal examples of G2GML are shown in Section \ref{subsec:minimal-examples}.

An RDF graph pattern in G2GML specifies the resources that should be mapped to PG components. Variables in RDF graph patterns are embedded into the corresponding location in the preceding PG patterns, which yield resultant PGs. 

A node map is described as a pair of a PG node pattern and an RDF graph pattern. Here, we define the syntax of the PG node pattern as follows. 
\begin{defi}[EBNF notation of PG node pattern]
\label{defi:pgnode}
\leavevmode
\begin{verbatim}
 node_pat  ::= "(" NODE  ":" LABEL  ("{" prop_list "}")? ")"
 prop_list ::= property "," prop_list | property
 property  ::= PROP_NAME ":" PROP_VAL
\end{verbatim}
\end{defi}

In Definition~\ref{defi:pgnode}, \texttt{NODE}, \texttt{LABEL}, \texttt{PROP\_NAME}, and \texttt{PROP\_VAL} are all identifiers.
A PG node pattern specifies a node, a label, and properties to be generated by the map. The \texttt{NODE} symbol specifies the ID of the generated node, which is replaced by the resource retrieved from the RDF graph. The \texttt{LABEL} symbol specifies the label of the resultant node. A label also serves as the name of the node map, which is referred from edge maps.
Therefore, the identifier of each label must be unique in a G2GML description.
Each node pattern contains zero or more properties in a \texttt{prop\_list}.
A \texttt{property} is a pair of \texttt{PROP\_NAME} and \texttt{PROP\_VAL} that describes the properties of nodes.
The identifiers \texttt{NODE} and \texttt{PROP\_VAL} must be names of variables in the succeeding RDF graph pattern, while \texttt{LABEL} and \texttt{PROP\_NAME} are regarded as text literals.
 
In contrast, a PG edge pattern specifies the pattern of the edge itself, its label, and properties, as well as the source and destination nodes.
Here, we define the syntax of the PG edge pattern as follows.
\begin{defi}[EBNF notation of PG edge pattern]
\label{defi:pgedge}
\leavevmode
\begin{verbatim}
edge_pat ::= "(" SRC  ":" SRC_LAB ") -" 
             "[" EDGE?  ":" EDGE_LAB  ("{" prop_list "}")? "]" 
             ("->" | "-") "(" DST ":" DST_LAB ")"
\end{verbatim}
\end{defi}

In Definition~\ref{defi:pgedge}, \texttt{SRC}, \texttt{SRC\_LAB}, \texttt{EDGE}, \texttt{EDGE\_LAB}, \texttt{DST}, and \texttt{DST\_LAB} are all identifiers. Identifiers of \texttt{SRC} and \texttt{DST} specify the variables in the succeeding RDF graph pattern that should be mapped to the endpoint nodes of resultant edges.
A \texttt{SRC\_LAB} and a \texttt{DST\_LAB} serve not only as labels of resultant nodes, but also as implicit constraints of the edge map. The resultant source and destination nodes must match patterns in the corresponding node maps (described in Section \ref{subsec:mapping-details} with a concrete example).
For this reason, \texttt{SRC\_LAB} and \texttt{DST\_LAB} must be defined in other node maps.
\texttt{EDGE}, \texttt{EDGE\_LAB}, and \texttt{prop\_list} can be described in the same manner as their counterparts in node maps.

The resultant edges can be either directed or undirected. If '\texttt{->}' is used as a delimiter between an edge and a destination part, the edge will be  directed.

In node and edge maps, a variable bound to \texttt{PROP\_VAL} may have multiple candidates for a single node or edge. If multiple candidates exist, they are concatenated into an array in the same manner as GROUP\_CONCAT in SPARQL. If \texttt{EDGE} is omitted in an edge map, such candidates are concatenated within groups of tuples of \texttt{SRC} and \texttt{DST}.

\subsection{Property Graph Model}
We define the PG model independent of specific graph database implementations. For the purpose of interoperability, we incorporate differences in PG models, taking into consideration multiple labels or property values for nodes and edges, as well as mixed graphs with both directed and undirected edges. The PG model that is redefined here requires the following characteristics:

\begin{itemize}
    \item A PG contains nodes and edges.
    \item Each node and edge can have zero or more labels.
    \item Each node and edge can have properties (key-value pairs).
    \item Each property can have multiple values.
    \item Each edge can be directed or undirected.
\end{itemize}
Formally, we define the PG model as follows.

\begin{defi}[Property Graph Model]
\leavevmode \vspace{1mm} \\
A \emph{property graph} is a tuple
$PG = \langle N, E_d, E_u, e, l_n, l_e, p_n, p_e\rangle$, where:
\begin{enumerate}
    \item $N$ is a set of nodes.
    \item $E_d$ is a set of directed edges.
    \item $E_u$ is a set of undirected edges.
    \item $e: E \to \langle N \times N \rangle$ is a function that yields the endpoints of each directed or undirected edge where $E := E_d \cup E_u$. If the edge is directed, the first node is the source, and the second node is the destination.
    \item $l_n : N \to 2^S$ is a function mapping each node to its multiple labels where $S$ is a set of all strings.
    \item $l_e : E \to 2^S$ is a function mapping each edge to its multiple labels.
    \item $p_n : N \to 2^P$ is a function used to assign nodes to their multiple properties. $P$ is a set of properties. Each property assumes the form $p = \langle k,v \rangle$, where $k \in S$ and $v \in 2^V$. Here $V$ is a set of values of arbitrary data types.
    \item $p_e : E \to 2^P$ is a function used to assign edges to their multiple properties.
\end{enumerate}
\end{defi}

\subsection{Serialization of Property Graphs}
According to our definition of the PG model, we propose serialization in flat text and JSON. The flat text format (PG format) performs better in terms of human readability and line-oriented processing, while the JSON format (JSON-PG format) is best used for server--client communication.

\begin{figure}
\center
\includegraphics[width=0.9\textwidth]{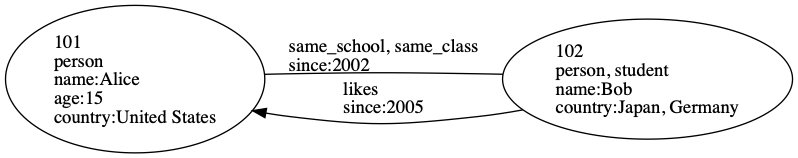}
\caption{Example of PG format (Visualization)}
\label{fig:pg_example_vis}
\end{figure}

The PG format has the following characteristics. An example is given in Figure~\ref{fig:example-pg}, which is visualized in Figure~\ref{fig:pg_example_vis}.

\begin{itemize}
    \item Each line describes a node or an edge.
    \item All elements in a line are separated by spaces or tabs.
    \item The first column of a node line contains the node ID.
    \item The first three columns of an edge line contain the source node ID, direction, and destination node ID.
    \item Each line can contain an arbitrary number of labels.
    \item Each line can contain an arbitrary number of properties (key-value pairs).
\end{itemize}

More formally, we describe the PG format in the EBNF notation as follows.
\vspace{1ex}
\begin{defi}[EBNF notation of the PG format]
\leavevmode
\begin{verbatim}
 pg         ::= (node | edge)+
 node       ::= NODE_ID labels properties NEWLINE
 edge       ::= NODE_ID direction NODE_ID labels properties NEWLINE
 labels     ::= label*
 properties ::= property*
 label      ::= ":" STRING
 property   ::= STRING ":" VALUE
 direction  ::= "--" | "->"
\end{verbatim}
\end{defi}

According to this definition, each property value of PGs is a set of items of any datatype. Meanwhile, our G2G mapping prototype implementation currently supports the three main datatypes (integer, double, and string) as property values, and those types are inferred from the format of serialized values.

Furthermore, we implemented command-line tools to convert formats between the flat PG and JSON-PG, where the latter follows the JSON syntax in addition to the above definition. We further transform them into formats for well-known graph databases such as Neo4j, Oracle Database, and Amazon Neptune. The practical use cases of our tools demonstrate that the proposed data model and formats have the capability to describe PG data used in existing graph databases (see \url{https://github.com/g2glab/pg}).

\begin{figure}[!t]
\begin{scriptsize}
\begin{verbatim}
# NODES
101  :person  name:Alice  age:15  country:"United States"
102  :person  :student  name:Bob  country:Japan  country:Germany

# EDGES
101 -- 102  :same_school  :same_class  since:2002
102 -> 101  :likes  since:2005
\end{verbatim}
\end{scriptsize}
\caption{Example of PG format}
\label{fig:example-pg}
\end{figure}

\subsection{Minimal Example}
\label{subsec:minimal-examples}
Figure~\ref{fig:example-g2g} shows the minimal example of G2G mapping from RDF data (Figure~\ref{fig:example-rdf}) to PG data (Figure~\ref{fig:output-pg}), representing the following five types of typical mapping.

\begin{itemize}
    \item Resource to node: In lines 2--4, the RDF resources with type \texttt{:Person} are mapped into the PG nodes using their IRIs as node IDs.
    \item Datatype property to node property: In lines 2--4, the RDF datatype property \texttt{:name} is mapped onto the PG node property key \texttt{name}. The literal objects \texttt{'Alice'} and \texttt{'Bob'} are mapped onto the node property values.
    \item Object property to edge: In lines 5--6, the RDF object property \texttt{:supervised\_by} is mapped onto the PG edge \texttt{supervised\_by}.
    \item Resource to edge: In lines 7--12, the RDF resource with type \texttt{:Email} is mapped onto the PG edge \texttt{emailed}. 
    \item Datatype property to edge property: In lines 7--12, the RDF datatype property \texttt{:year} and \texttt{:attachment} are mapped onto the PG edge property \texttt{year} and \texttt{attachment}. The literal objects \texttt{2017} and \texttt{'01.pdf'} are mapped onto the edge property values.
\end{itemize}

\begin{figure}[!t]
\begin{scriptsize}
\begin{verbatim}
@prefix : <http://example.org/> .
:person1 a :Person ;
         :name 'Alice' .
:person2 a :Person ;
         :name 'Bob' .
:person1 :supervised_by :person2 .
[] a :Email ;
   :sender     :person1 ;
   :receiver   :person2 ;
   :year       2017 ;
   :attachment '01.pdf' .
\end{verbatim}
\end{scriptsize}
\caption{Example of input RDF data}
\label{fig:example-rdf}
\end{figure}

\begin{figure}[!t]
\begin{scriptsize}
\begin{verbatim}
PREFIX : <http://example.org/>
(p:person {name:n})
    ?p a :Person .
    ?p :name ?n .
(p1:person)-[:supervised_by]->(p2:person)
    ?p1 :supervised_by ?p2 .
(p1:person)-[:emailed {year:y, attachment:a}]->(p2:person)
    ?f a :Email ;
       :sender   ?p1 ;
       :receiver ?p2 ;
       :year     ?y .
    OPTIONAL { ?f :attachment ?a }
\end{verbatim}
\end{scriptsize}
\caption{Example of G2G mapping definition}
\label{fig:example-g2g}
\end{figure}

\begin{figure}[!t]
\begin{scriptsize}
\begin{verbatim}
"http://example.org/person1" :person name:Alice
"http://example.org/person2" :person name:Bob
"http://example.org/person1" -> "http://example.org/person2" :supervised_by
"http://example.org/person1" -> "http://example.org/person2" :emailed year:2017 attachment:"01.pdf"
\end{verbatim}
\end{scriptsize}
\caption{Example of output PG data}
\label{fig:output-pg}
\end{figure}

\subsection{Mapping Details}
\label{subsec:mapping-details}
The G2G mapping based on G2GML is designed to be intuitive in general as shown above; however, there are several discussions regarding the details of mapping.

\subsubsection{Referencing Node Labels}
The node labels in edge maps must be defined in node maps (described in Section 2.2) to reference the conditions. This means that the conditions for defining \texttt{person} are imported into the definition of \texttt{supervised\_by} and \texttt{emailed} relationships. Therefore, the \texttt{supervised\_by} relationship will be generated only between the nodes satisfying the \texttt{person} condition. In this manner, users can retrieve the PG datasets that are intuitively consistent (it is also possible to avoid specifying node labels). In contrast, edge conditions are not inherited to nodes, such that the nodes are retrieved without all relationships, \texttt{supervised\_by} and \texttt{emailed}. Although the nodes with no relationship also satisfy the conditions, the prototype implementation does not retrieve such orphan nodes by default from a practical perspective.

\subsubsection{Multi-Edges}
When a pair of RDF resources has multiple relationships, those will be converted to multiple PG edges instead of a single PG edge with multiple properties. The G2GML in Figure~\ref{fig:example-g2g} generates two PG edges for the RDF data in Figure~\ref{fig:example-rdf2} to maintain the information that Alice emailed Bob twice (Figure~\ref{fig:example-pg2}).

\begin{figure}[!t]
\begin{scriptsize}
\begin{verbatim}
@prefix : <http://example.org/> .
:person1 a :Person ;
         :name 'Alice' .
:person2 a :Person ;
         :name 'Bob' .
[] a :Email ;
   :sender   :person1 ;
   :receiver :person2 ;
   :year     2017 .
[] a :Email ;
   :sender   :person1 ;
   :receiver :person2 ;
   :year     2018 .
\end{verbatim}
\end{scriptsize}
\caption{Example of input RDF data (multi-edges)}
\label{fig:example-rdf2}
\end{figure}

\begin{figure}[!t]
\begin{scriptsize}
\begin{verbatim}
"http://example.org/person1" :person name:Alice
"http://example.org/person2" :person name:Bob
"http://example.org/person1" -> "http://example.org/person2" :emailed year:2017
"http://example.org/person1" -> "http://example.org/person2" :emailed year:2018
\end{verbatim}
\end{scriptsize}
\caption{Example of output PG data (multi-edges)}
\label{fig:example-pg2}
\end{figure}

\subsubsection{List of Property Values}
Each PG property can assume multiple values. In an RDF that includes the same data property multiple times, the values are assumed to be the members of a list. The G2GML in Figure~\ref{fig:example-g2g} generates one PG edge with two properties for the RDF data in Figure~\ref{fig:example-rdf3}, keeping the information that Alice emailed Bob once, and the email contained two attachments (Figure~\ref{fig:example-pg3}).

\begin{figure}[!t]
\begin{scriptsize}
\begin{verbatim}

@prefix : <http://example.org/> .
:person1 a :Person ;
         :name 'Alice' .
:person2 a :Person ;
         :name 'Bob' .
[] a :Email ;
   :sender     :person1 ;
   :receiver   :person2 ;
   :year       2017 ;
   :attachment '01.pdf' ;
   :attachment '02.pdf' .

\end{verbatim}
\end{scriptsize}
\caption{Example of input RDF data (list of property values)}
\label{fig:example-rdf3}
\end{figure}

\begin{figure}[!t]
\begin{scriptsize}
\begin{verbatim}

"http://example.org/person1" :person name:Alice
"http://example.org/person2" :person name:Bob
"http://example.org/person1" -> "http://example.org/person2" :emailed year:2017 attachment:"01.pdf" attachment:"02.pdf"

\end{verbatim}
\end{scriptsize}
\caption{Example of output PG data (list of property values)}
\label{fig:example-pg3}
\end{figure}

\section{Use Cases}
 
We present several use cases to show how to apply the G2G mapping framework, where we utilize publicly available RDF datasets through SPARQL endpoints.

\subsection{Using Wikidata}
 
\begin{figure}
\center
\includegraphics[width=0.7\textwidth]{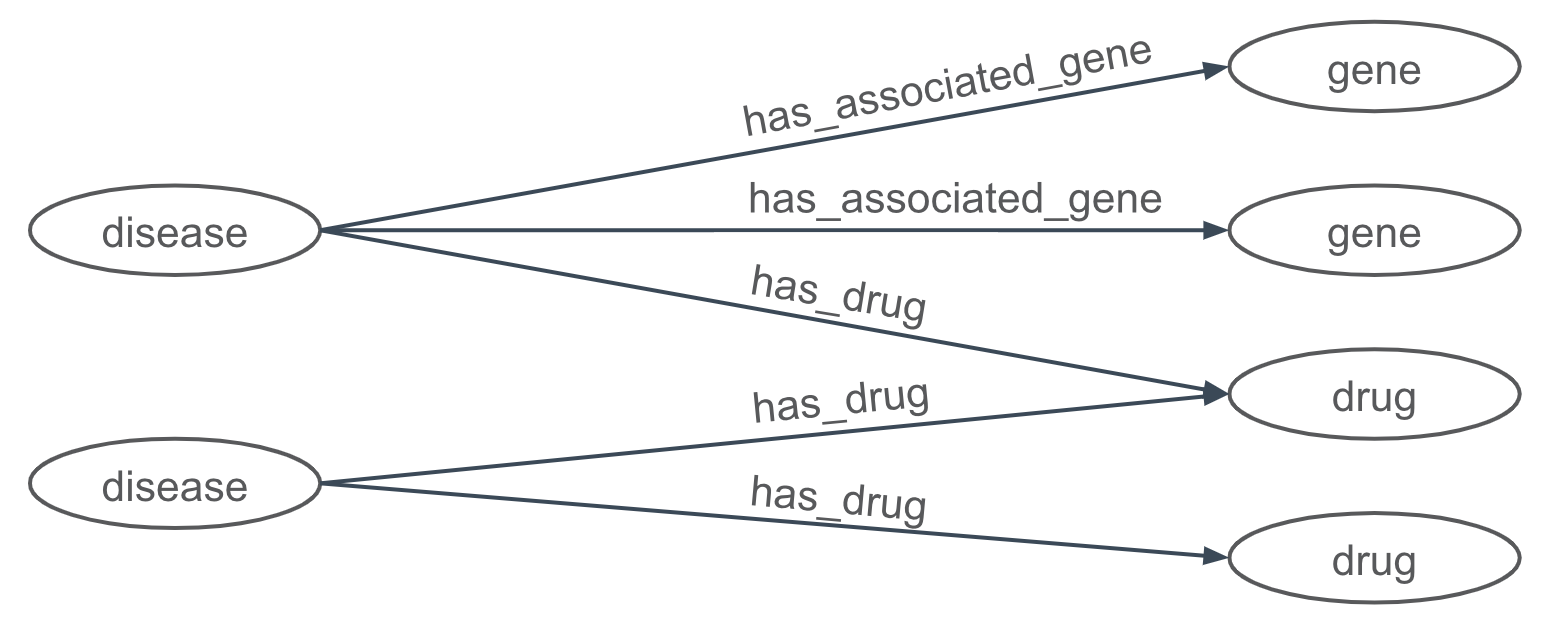}
\caption{Schematic relations of Wikidata entities}
\label{fig:wikidata_schema.png}
\end{figure}

Wikidata is a useful source of open data from various domains.
We present an example of mapping using a disease dataset from Wikidata, 
where each disease is associated with zero or more genes and drugs.
Figure~\ref{fig:wikidata_schema.png} illustrates the schematic relationships between those entities. 
This subset of Wikidata can be converted as a PG using the G2G mapping framework by writing a mapping file in G2GML (see Figure~\ref{fig:g2gml_wikidata}).
Here, we focus on human genes and specify the necessary conditions in G2GML.
Each instance of \texttt{Q12136} (disease) can have a property \texttt{P2176} (drug used for treatment), which is thereby linked to items of \texttt{Q12140} (medication). Further, each disease can have a property \texttt{P2293} (genetic association), which is thereby linked to items of \texttt{Q7186} (gene).
The resultant PG includes 4696 diseases, 4496 human genes, and 1287 drugs. The total numbers of nodes and edges are summarized in Table~\ref{table:numbers}. As a result of G2G mapping, the numbers of nodes and edges are reduced to almost half and one third, respectively.
 
\begin{figure}[!t]
\vspace{2mm}
\begin{scriptsize}
\begin{verbatim}
 
PREFIX wd: <http://www.wikidata.org/entity/>
PREFIX wdt: <http://www.wikidata.org/prop/direct/>
 
(g:human_gene {symbol: s})
    ?g wdt:P31 wd:Q7187 ;      # "instance of" "gene"
       wdt:P703 wd:Q15978631 ; # "found in taxon" "Homo sapiens"
       wdt:P353 ?s .           # "HGNC gene symbol"
 
(d:disease {name: n})
    ?d wdt:P31 wd:Q12136 ;     # "instance of" "disease"
       rdfs:label ?l .
    FILTER(lang(?l) = "en")
    BIND(str(?l) AS ?n)
 
(m:drug {name: n})
    ?m wdt:P31 wd:Q12140 ;     # "instance of" "medication"
       rdfs:label ?l .
    FILTER(lang(?l) = "en")
    BIND(str(?l) AS ?n)
 
(d:disease)-[:has_associated_gene]->(g:human_gene)
    ?d wdt:P2293 ?g .          # "genetic association"
 
(d:disease)-[:has_drug]->(m:drug)
    ?d wdt:P2176 ?m .          # "drug used for treatment"
 
\end{verbatim}
\end{scriptsize}
\caption{G2GML for Wikidata mapping}
\label{fig:g2gml_wikidata}
\end{figure}
 
\subsection{Using DBpedia}
 
\begin{figure}
\center
\includegraphics[width=1.0\textwidth]{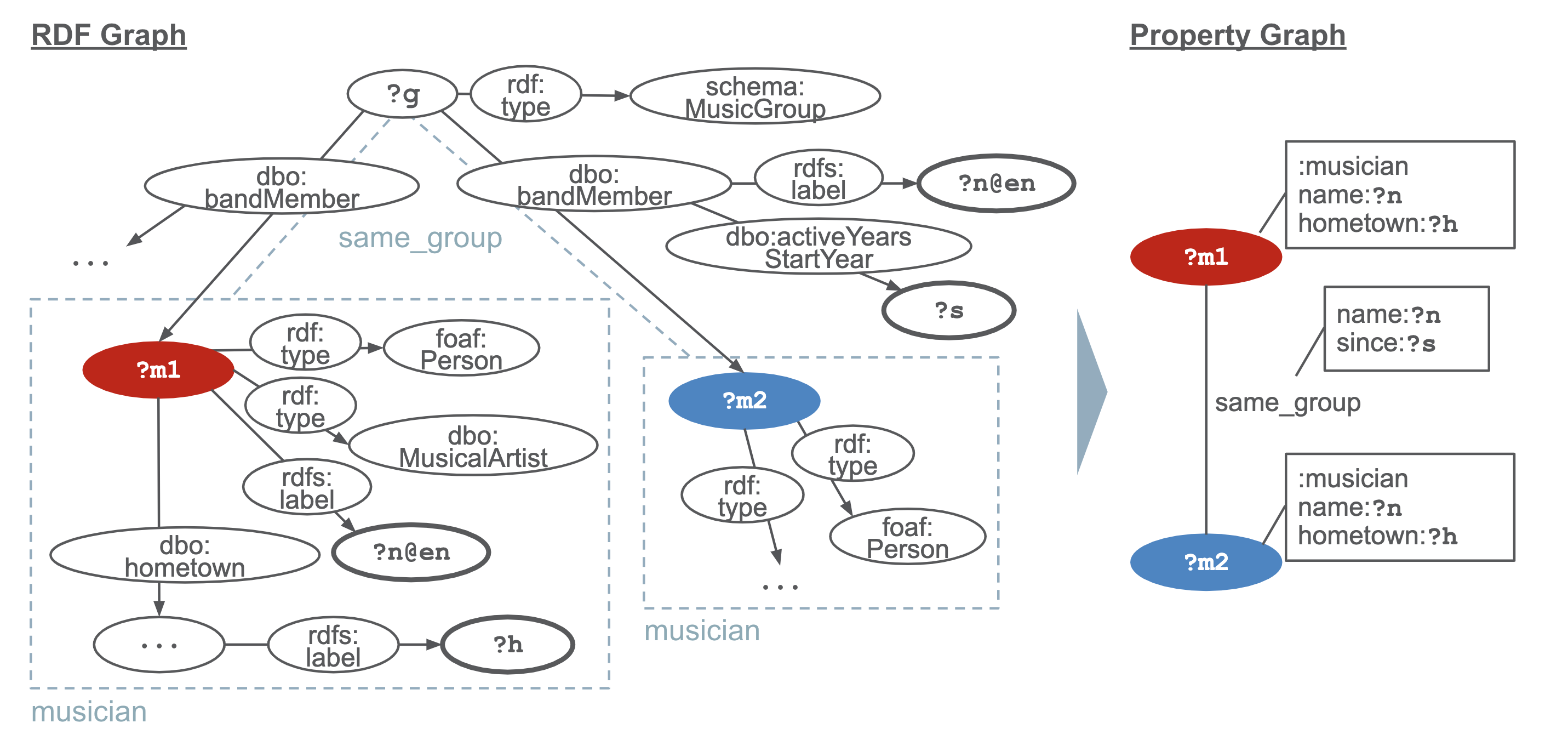}
\caption{Schematic example of DBpedia mapping}
\label{fig:conversion}
\end{figure}

Figure~\ref{fig:conversion} schematically illustrates an example of the G2G mapping to convert RDF data retrieved from DBpedia into PG data. 
Focusing on a relationship where two musicians (\texttt{?m1} and \texttt{?m2}) belong to the same group, this information can be represented in the PG shown on the right side of the figure. The relationships are originally presented as independent resources \texttt{?g}. In the mapping, we map \texttt{?g}, \texttt{?n}, and \texttt{?s} onto a PG edge labeled \texttt{same\_group} with attributes. Such compaction by mapping is useful in numerous use cases, for example, when users are interested only in relationships between musicians.
 
The mapping rules for this example are specified in the G2GML description shown in Figure~\ref{fig:g2gml}. The G2GML description contains node mapping for \texttt{musician} entities and edge mapping for \texttt{same\_group} relationships.
In each specified PG pattern, \texttt{\{m, n, h\}} and \texttt{\{m1, m2, n, s\}} are used as variables to reconstruct resources and literals extracted from RDF graphs.

\begin{figure}[!t]
\vspace{2mm}
\begin{scriptsize}
\begin{verbatim}
PREFIX rdf: <http://www.w3.org/1999/02/22-rdf-syntax-ns#>
PREFIX rdfs: <http://www.w3.org/2000/01/rdf-schema#>
PREFIX schema: <http://schema.org/>
PREFIX dbo: <http://dbpedia.org/ontology/>
PREFIX foaf: <http://xmlns.com/foaf/0.1/>
 
# Node mapping
(m:musician {name:n, hometown:h})                            # PG Pattern
    ?m rdf:type foaf:Person , dbo:MusicalArtist ;            # RDF Pattern
       rdfs:label ?n . FILTER(lang(?n) = "en") .
    OPTIONAL { ?m dbo:hometown/rdfs:label ?h . FILTER(lang(?h) = "en") }
 
# Edge mapping
(m1:musician)-[:same_group {name:n, since:s}]-(m2:musician)   # PG Pattern
    ?g rdf:type schema:MusicGroup ;                          # RDF Pattern
       dbo:bandMember ?m1 , ?m2 . FILTER(?m1 != ?m2)
    OPTIONAL { ?g rdfs:label ?n . FILTER(lang(?n) = "en")}
    OPTIONAL { ?g dbo:activeYearsStartYear ?s }
\end{verbatim}
\end{scriptsize}
\caption{G2GML for DBpedia mapping}
\label{fig:g2gml}
\end{figure}
 
\begin{figure}[!t]
\vspace{2mm}
\begin{scriptsize}
\begin{verbatim}
# SPARQL
PREFIX rdf: <http://www.w3.org/1999/02/22-rdf-syntax-ns#>
PREFIX rdfs: <http://www.w3.org/2000/01/rdf-schema#>
PREFIX schema: <http://schema.org/>
PREFIX dbo: <http://dbpedia.org/ontology/>
PREFIX foaf: <http://xmlns.com/foaf/0.1/>
 
SELECT DISTINCT ?nam1 ?nam2
WHERE {
    ?mus1 rdf:type foaf:Person , dbo:MusicalArtist .
    ?mus2 rdf:type foaf:Person , dbo:MusicalArtist .
    ?mus1 rdfs:label ?nam1 . FILTER(lang(?nam1) = "ja") .
    ?mus1 rdfs:label ?nam2 . FILTER(lang(?nam2) = "ja") .
    ?grp a schema:MusicGroup ;
         dbo:bandMember ?mus1 , ?mus2 .
    FILTER(?mus1 != ?mus2)
}

# Cypher
MATCH (m1)-[:same_group]-(m2) RETURN DISTINCT m1.name, m2.name

# PGQL
SELECT DISTINCT m1.name, m2.name MATCH (m1)-[:same_group]-(m2)
\end{verbatim}
\end{scriptsize}
\caption{SPARQL, Cypher, and PGQL}
\label{fig:sparql}
\end{figure}
 
Figure~\ref{fig:sparql} shows queries to retrieve the pairs of musicians that are in the same group in SPARQL and graph query languages ~\cite{openCypher}~\cite{pgql}. The queries in Cypher and PGQL are more succinct owing to the simple structure of the PG obtained by G2G mapping.

\begin{table}[h]
    \centering
    \begin{tabular}{l|r|r|r|r}
        \hline
        & RDF nodes & RDF edges & PG nodes & PG edges \\
        \hline
        Wikidata disease & 20,692 & 36,826 & 10,477 & 11,770 \\
        DBpedia musician & 23,846 & 32,808 & 7,069 & 10,755 \\
        \hline
    \end{tabular}
    \caption{Number of nodes and edges in use cases}
    \label{table:numbers}
\end{table}

\section{Availability}
The prototype implementation of G2G mapping is written in JavaScript and can be executed using Node.js in the command line. It has an endpoint mode and a local file mode. The local file mode uses Apache Jena ARQ to execute SPARQL queries internally, whereas the endpoint mode accesses SPARQL endpoints via the Internet. An example of the usage in the endpoint mode is as follows:

\texttt{\$ g2g musician.g2g http://dbpedia.org/sparql}

\noindent where the first argument is a G2GML description file, and the second argument is the target SPARQL endpoint, which provides the source RDF dataset.

\begin{figure}
\center
\includegraphics[width=1.0\textwidth]{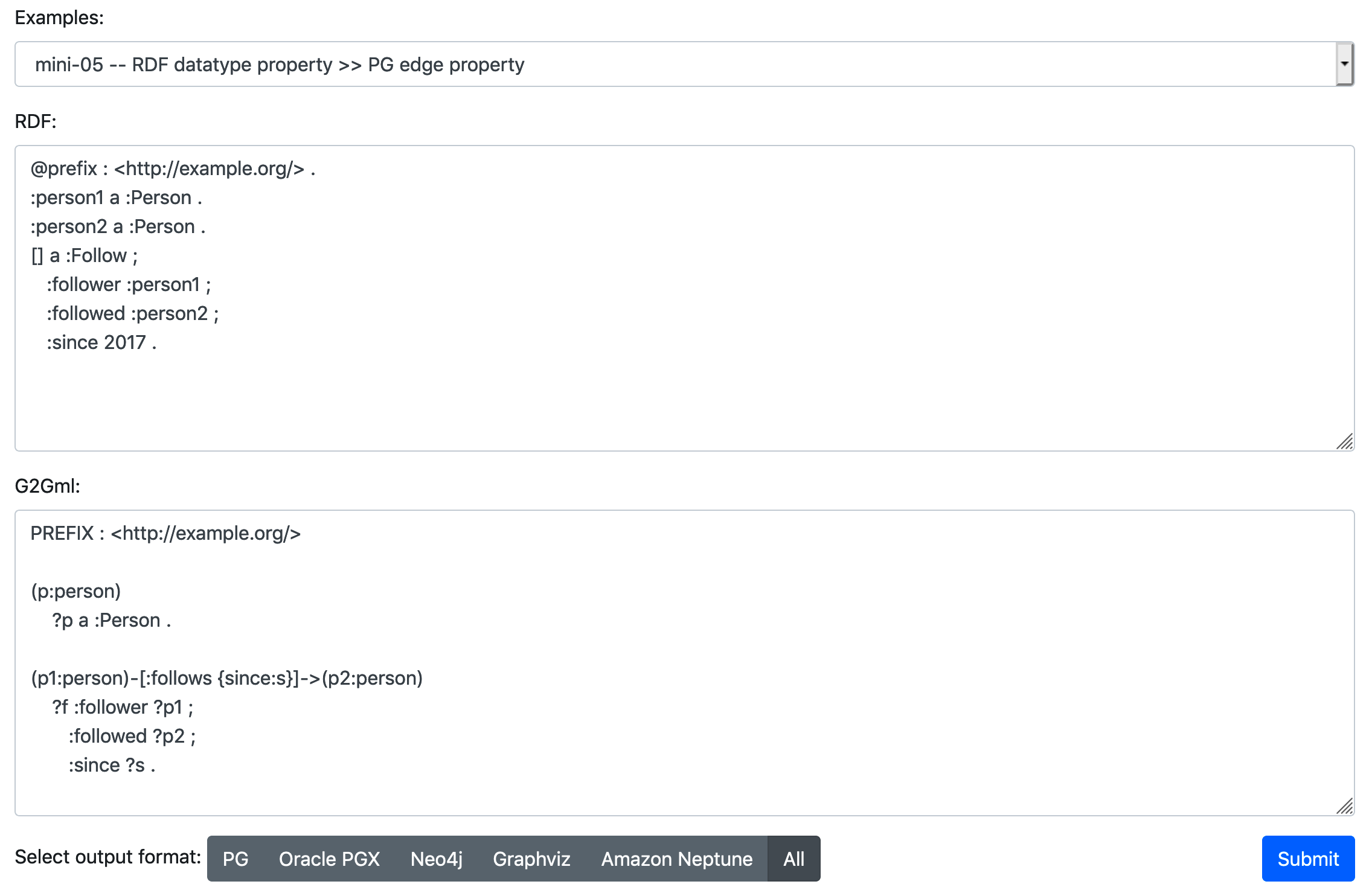}
\caption{G2G mapping demonstration site}
\label{fig:sandbox}
\end{figure}

Furthermore, a Docker image (\url{https://hub.docker.com/r/g2glab/g2g}) and a demonstration site of the G2G mapping framework (\url{https://purl.org/g2gml}) are available (Figure~\ref{fig:sandbox}).

In future adoptions of this framework, we expect two scenarios: The use of the standalone G2G mapping tools, such as our implementation, for generating PG datasets from RDF resources. Further, the adoption of the framework within the database management systems, which support both RDF and PG datasets.

\section{Related Work}

\subsection{Property Graph Model and Serialization}
Recently, graph data has increasingly attracted attention, leading to a plethora of database implementations. 
Thus far, different data models and query languages have been available for these graph database implementations.
Consequently, there have been community-based activities for standardizing the graph query language for interoperable use of PG databases~\cite{angles3}. Similarly, the standardization of the PG model for various database implementations enhances the interoperable use of graph data. There is indeed a demand for graph standardization in the community, which was recently discussed in a W3C workshop~\cite{w3c}. Another proposal for the PG model and serialization~\cite{tomaszuk} that was similar to ours was presented in the workshop. Notably, that the two independent studies converged to a similar solution. They seem to be interchangeable; however, this remains to be tested. Future studies should address collaboration and standardization of the data model.
In particular, our serialization has implementations for some of the major database engines and has the potential to further cover various database engines. 
The serialization formats that are independent of specific database implementations will increase the interoperability of graph databases and make it easier for users to import accumulated graph data.

\subsection{Graph to Graph Mapping}
A preceding study on converting existing data into graph data included an effort to convert relational databases into graph databases~\cite{virgilio1}. 
However, given that RDF has prevailed as a standardized data model in scientific communities, considering mapping based on the RDF model is crucial. The interoperability of RDF and PG~\cite{hartig,angles4,das,thakkar} has been discussed, and efforts were made to develop methods to convert RDF into PG~\cite{tomaszuk1,virgilio}. However, considering the flexibility regarding the type of information that can be expressed by edges in property graphs, a novel method for controlling the mapping is necessary.
We discuss the comparison of controlled mapping and direct mapping later in this section.

To the best of our knowledge, this study presents the first attempt to develop a framework for controlled mapping between RDF and PG. 
Notably, the designed G2GML is a declarative mapping language. 
As a merit of the declarative description, we can concentrate on the core logic of mappings. In the sense that the mapping process generates new graph data on the basis of existing graph data, it has a close relation to the semantic inference. A similar concept is found in the SPARQL CONSTRUCT queries. While the SPARQL CONSTRUCT clause defines mapping on the same data model, G2GML defines mapping between different data models. 
Thus, G2GML is considered as a specific extension of the SPARQL CONSTRUCT clause for generation or inference of PG data.

A previous study compared the performance of RDF triple stores, such as Blazegraph and Virtuoso, and Neo4j~\cite{alocci}. They concluded that RDF has an advantage in terms of performance and interoperability. However, they tested only Neo4j as an implementation of the PG model. It is necessary to update such benchmarks using additional implementations, which requires improved interoperability and standardization of the PG model. Our study is expected to contribute to the improved interoperability of PG models, although a continuous discussion is necessary for the standardization of query languages for the PG model to achieve an interoperable ecosystem of both RDF and PG, along with a fair benchmarking.

\subsubsection{Direct Mapping}

Other mapping frameworks, such as Neosemantics (a Neo4j plugins), propose a method to convert RDF datasets without mapping definitions (for convenience, we call such methods direct mapping). However, the following capabilities are essential in practical usage.

1. Filtering data - RDF is designed for the web of data. When the source RDF dataset is retrieved from the public web space, it is inefficient (or even unrealistic) to convert the whole connected dataset. In G2GML, users can specify the resources they need, such that the SPARQL endpoints can return the filtered dataset. (This design is described in Section 2.2.)

2. Mapping details - There is no common ruleset to uniquely map RDF terms to PG elements (labels, key-value pairs of properties) and to name new PG elements. Further, the conversion rules of multi-edges and lists of property values (discussed in Section 2.6) are not always obvious to users. Therefore, defining a method for mapping details is necessary to create precise data in actual use cases.

3. Schema definition - We assume that mapping is often used for developing specific applications on top of PG datasets. In this development, the schema of the dataset should be known, while the original RDF data source could contain more information that is not covered by the schema. G2GML helps developers understand the data schema in its intuitive definition (separately for nodes and edges, and their referencing - discussed in Section 2.6), while direct mapping has the potential to generate PGs without defining a schema.

We observe a similar discussion in the conversion from the relational model to RDF, where are two W3C standards, i.e., Direct Mapping~\cite{dm} and R2RML~\cite{r2rml}.

\section{Conclusion}

We designed the G2GML for mapping RDF graphs to PGs and developed its prototype implementation. To ensure a clear definition of this mapping, we defined the PG model independent of specific graph database implementations and proposed its exchangeable serialization formats.

The advantage of using RDF is that different applications can retrieve necessary information from the same integrated datasets owing to the interoperable nature of semantic modeling.

For such increasing RDF resources, graph databases are potentially the ideal data storage as the property graph model can naturally preserve the relationship information semantically defined in RDF. G2GML therefore plays an important role in this data transformation process. 

Various graph database implementations are actively developed and the standardization of query languages is currently ongoing. We expect the G2GML or its enhanced mapping framework to be supported by database management systems and other software. We believe that our efforts of generalization and prototype implementation will promote further discussion.

\section*{Acknowledgements}
Part of the work was conducted in BioHackathon meetings (\url{http://www.biohackathon.org}). We thank Yuka Tsujii for helping create the figures. We thank Ramona Röß for careful review of the manuscript and useful comments.

%
%
%
\bibliographystyle{splncs04}
%

\end{document}